# Spinning Janus doublets driven in uniform AC electric fields

Alicia Boymelgreen[1], Gilad Yossifon[1*], Sinwook Park[1], Touvia Miloh[2]

[1]*Faculty of Mechanical Engineering, Micro- and Nanofluidics Laboratory, Technion – Israel Institute of Technology, Haifa 32000, Israel*

[2]*School of Mechanical Engineering, University of Tel-Aviv, Tel-Aviv 69978, Israel*

We provide an experimental proof of concept for a robust, continuously rotating microstructure - consisting of two metallodielectric (gold-polystyrene) Janus particles rigidly attached to each other – which is driven in uniform ac fields by asymmetric induced-charge electroosmosis. The pairs (doublets) are stabilized on the substrate surface which is parallel to the plane of view and normal to the direction of the applied electric field. We find that the radius of orbit and angular velocity of the pair are predominantly dependent on the relative orientations of the interfaces between the metallic and dielectric hemispheres and that the electrohydrodynamic particle-particle interactions are small. Additionally, we verify that both the angular and linear velocities of the pair are proportional to the square of the applied field which is consistent with the theory for non-linear electrokinetics. A simple kinematic rigid body model is used to predict the paths and doublet velocities (angular and linear) based on their relative orientations with good agreement.

PACS numbers: 47.57.jd, 47.57.J, 47.61.-k

The influence of broken symmetries and non-linear responses on the motion of matter in external fields is a topic of fundamental interest. In uniform AC fields, since the time averaged forcing is zero, even exotic fixed surface charge distributions (e.g., [1,2]) cannot result in continuous rotation, while the non-linear effects of dielectrophoresis (DEP) and induced-charge electrophoresis (ICEP), require some asymmetry to be introduced into the system to generate even linear translation.

Perhaps the most well-known method of inducing rotation is via the use of rotating fields, both electric and magnetic, where both single (e.g., [3,4]) and chains (e.g., [5–7]) of polarizable particles have been shown to rotate as their induced dipoles align with the continuously changing field. In uniform fields, rotation-to-alignment stemming from asymmetric non-linear ICEO around non-spherical geometries was first observed by Murtsovkin [8] and later examined analytically and experimentally [9–14]. However, in the absence of a rotating field, such motion is transient; acting only until the induced dipole is aligned with the field direction.

Rather than focusing solely on geometric asymmetry, Squires and Bazant [10] suggested that net mobility may also be attained by particles with heterogeneous electric properties; effectively describing a metallodielectric Janus particle (JP) in which one hemisphere is conducting (infinitely polarizable) and the other dielectric. Since then, net translation due to DEP [15,16] and ICEP [17,18] as well as instantaneous flip-flop rotation [19] have been demonstrated experimentally for single JPs.

In a uniform field, and at low frequencies (i.e., between the DC (inclusive) and Maxwell-Wagner limits), ICEP dominates DEP [20] and the combination of a stronger induced-charge electroosmotic (ICEO) flow around the more polarizable hemisphere of the JP and alignment of the dipole within, causes the particle to rotate so that the interface between the two hemispheres aligns itself with the electric field [10,17]. The ICEO flow around the conducting hemisphere then acts as a "jet", propelling the particle forward, perpendicular to the field, in the direction of its dielectric end with a velocity of [21]

$$U_{ICEP} = \left(\frac{\varepsilon R E^2}{\eta}\right) \frac{9/32}{1 + 2(\lambda_0 \varepsilon_1/\varepsilon)^{-1} + 45/64}, \qquad (0)$$

where $E$ denotes the magnitude of the applied electric field, $\eta$ the dynamic viscosity of the solute and $R$ the particle radius. The parameter $\lambda_0$ represents the dimensionless thickness of the Debye layer, normalized by $R$ so that for a relatively large (micron size) particle, $\lambda_0 \to 0$. The permittivities of the electrolyte and the conducting hemisphere are denoted by $\varepsilon$ and $\varepsilon_1$ respectively.

Based on the propensity of JPs to move perpendicular to the electric field with their dielectric end facing forward - always reorienting themselves to this stable state - Squires and Bazant [10] predicted an "ever-rotating" structure, comprised of two or three rigidly attached Janus particles which would rotate *continuously* under a *uniform* electric field.

To the best of our knowledge, the current contribution, wherein we examine the mobility of doublets of two rigidly attached metallodielectric Janus spheres suspended in DI water, aside from constituting a first experimental proof-of-concept for the theoretical model of Squires and Bazant [10], also holds significance as a first general demonstration of continuous rotation in uniform AC fields driven by symmetry-breaking. It is noted however, that in contrast to the design of Squires and Bazant [10], here the field is applied perpendicular to the plane of rotation rather than parallel thereto, resulting in an angular velocity which is constant rather than a function of the angle of the structure to the applied field. Additionally, this configuration facilitates measurable control of the ratio of rotation to linear translation via the orientation of the interfaces between the two particles to each other.

It is noted that previously, similarly sustained rotation combined with translation has been observed for doublets or clusters of catalytic self-propelling particles spherical JPs [22,23] as well as by imposing additional asymmetry on single spherical particles [24,25] and manufacturing rod or rotor-like structures [26–28]. However, one of the challenges associated with such self-propelling "micromotors" [29] is the ability to control the direction of motion and to this end, the integration of external electric and magnetic fields into such systems is currently being investigated [29–32] . By using these fields to drive the particle motion itself, we can develop a system which is externally controlled, non-catalytic (fuel-free) and may be switched on and off at will with potential applications ranging from targeted delivery in lab on a chip systems [33] to bottom-up material manufacture and optical displays [34]. Additionally, in contrast to applied rotating fields, here both translation and rotation may be obtained within the same system, suggesting integration of such particles as portable micromixers or even rotary motors [10] in applications which already rely on uniform applied fields such as separation analysis systems.

The Janus particles were prepared following the methodology in [17,19]. Commercially available suspensions of fluorescent colloids $4.8\mu m$ in diameter (Fluoro-Max) were rinsed in DI water three times and then suspended in isopropanol (ISP) to enhance the uniformity of the monolayer in the next step [19] when the particles are spin coated on a glass substrate. The coated glass slide was then evaporated with 10nm chrome for adhesion, followed by 50nm gold. The monolayer was broken up and the Janus particles were re-suspended in DI water by sonication.

The JP solution was inserted inside a silicone reservoir of depth 2mm and diameter 4.5mm (Grace Bio). The reservoirs were sandwiched between two glass substrates coated with Indium Tin Oxide (ITO) (Delta Technologies Ltd) which served as the electrodes ( Figure 1). They were connected to an Agilent signal generator (33250A) and a 50x amplifier (Falco Systems, WMA 300). The applied voltages both into the amplifier and the chip were monitored using an oscilloscope (Tektronix, TPS 2024). Based on the experimental observation of a decay in particle velocities at high frequencies [17], we chose to activate the system at a frequency of 1.5kHz which lies in the optimal range. The applied voltages were varied between 50-140Vp-p. Finally, the particle motion was observed using a Nikon TI inverted epifluorescent microscope and recorded with an Andor Neo sCMOS camera at a rate of 5fps. To obtain the velocity dispersion, particles were tracked in Image J software using the Speckle Tracker plugin [35] and the $x, y$ positions in each frame were recorded.

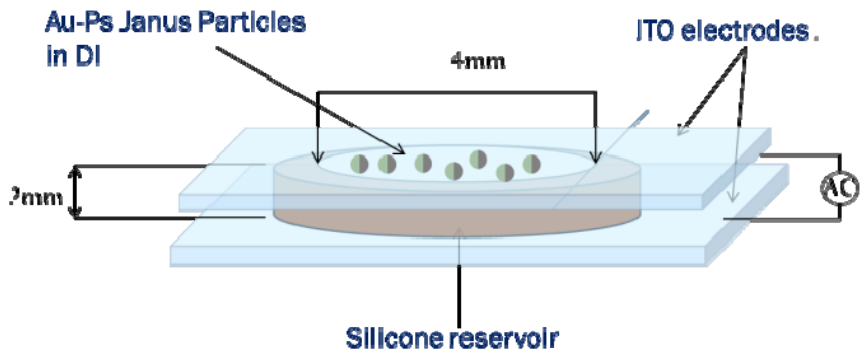

**Figure 1: Experimental setup consisting of a silicone reservoir between two parallel ITO covered glass slides. The uniform electric field is normal to the wall and plane of view**

In contrast to previous work on ICEP of Janus spheres [17,18], where the applied field is parallel to the channel wall, here we use a perpendicular field (Figure 1). As a result, the stability of the Janus spheres aligns their interface orthogonal to the wall which when combined with the wall attraction effect (see [17,36,37] although note that the direction of the applied field is not the same) results in 3D symmetry breaking and persistent motion parallel to the wall in the plane of view similar to that observed by Ebbens et al [38] for catalytically driven doublets and in contrast to the upwards translating doublets of Gangwal et al (supplementary materials in [17]).

The pairs (doublets) of JPs are formed spontaneously during the manufacturing process and are observed to act as a single rigid body with no relative motion between the spheres (see attached videos in supplementary materials [39]). In Figure 2, the pathlines of three such doublets are plotted alongside that of a single particle. A second set of three pairs has been provided in the supplementary material [39]. The applied voltage is held constant at $1.5 kHz$ and $76 V_{p-p}$, with an RMS field of $134\ V/cm$. In contrast to the single particle (Figure 2a), the three pairs in Figure 2b-d undergo marked rotation rather than simple linear translation, with significant variations in their radius of orbit and angular velocities. The latter quantity is represented by the number of points on a given orbit as all the points are equally spaced in time $(\Delta t = 0.2s)$. It is noted that since the stability analysis does not restrict particle motion in the lateral direction (parallel to the wall), the observation that single particles traveled in different directions with no apparent pattern indicates that there is no background hydrodynamic flow in the near wall region.

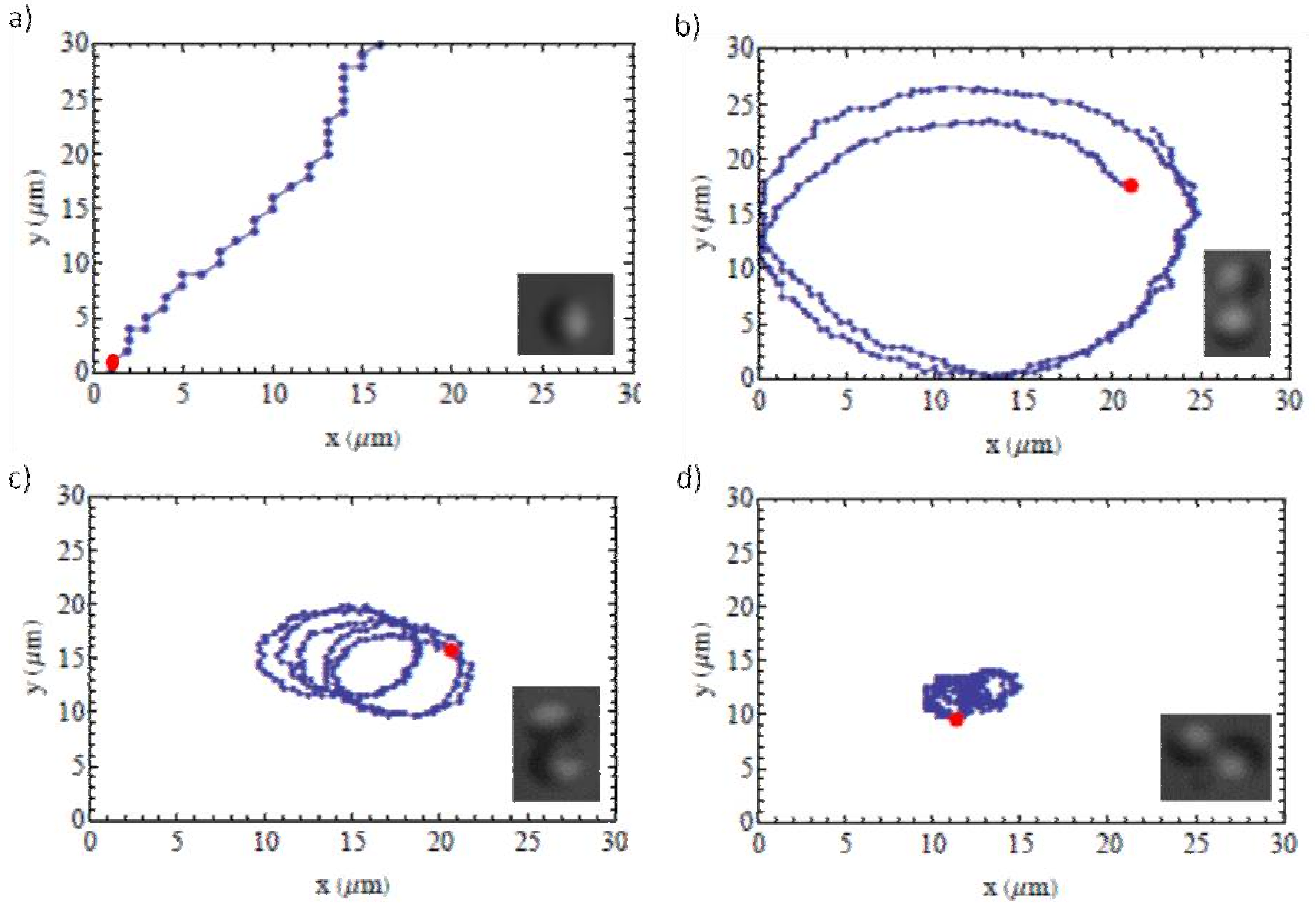

**Figure 2: Experimentally measured pathlines of (a) a single particle and (b-d) three pairs with different relative orientations of their metallo-dielectric interfaces. The red dot indicates the starting point of the trajectory. The gold-coated hemispheres appear darker.**

Although the proximity of the adjacent spheres suggests the presence of electrohydrodynamic particle-particle effects, we observe that the prevailing velocity and trajectory of the JP doublet are predominantly functions of the geometric interfacial alignment of the two particles and may be predicted using a simple kinematic rigid body model (Figure 3a). To this end, we approximate the center of rotation $C$ of the doublet trajectory to be located at the contact point between the two spheres which are joined by their line of centres, $r_{AB}$. Each sphere is presumed to have its own linear velocity vector, $\vec{V}_A, \vec{V}_B$, the magnitudes of which are equal to that of a single particle translating near and parallel to the wall (denoted by $V_{JP}$). The velocity direction is always normal to the Janus interface pointing into the dielectric hemisphere. We define the angles (measured in the anticlockwise direction) between the line-of-centres and the velocity vectors of the two particles as $\alpha$ and $\beta$ (see Fig 3a).

In the absence of particle-particle interactions, the rigid body translational velocity can be simply given as the vector average of the two velocities, i.e., $\vec{V}_C = \tfrac{1}{2}(\vec{V}_A + \vec{V}_B)$, where the magnitudes of $\vec{V}_A, \vec{V}_B$ are equal to that of a single translating particle $(V_{JP})$, resulting in $|\vec{V}_C| = \cos\left(\tfrac{1}{2}(\alpha - \beta)\right) V_{JP}$. The angular velocity around the $z$ axis (passing through $C$) is given as $\omega = (\sin\alpha - \sin\beta) V_{JP} / 2R$ where $R$ denotes the radius of a single particle. Finally, the radius of orbit, which is driven by the misalignment between the JP interfaces can be related to the translational velocity according to $r_{orbit} = |\vec{V}_C| / \omega$.

In Figures 3b-d, we compare a single complete cycle from the experimental trajectories of the three pairs illustrated in Figure 2b-d with those calculated from the rigid body kinematic model. Each line represents half the line-of-centers between the two spheres (marked as $r_{AC}$ in Figure 3a) with the dashed and solid lines representing the experimental results and kinematic model respectively. The lines are equally spaced in time $(\Delta t = 0.4s)$ so that again, the number of lines per cycle and their spacing correlate to the JP angular velocity.

The 'theoretical' angular and translation velocities $\omega_{KIN}$ and $V_{C-KIN}$, predicted by the kinematic model, are obtained by approximating $\alpha$ and $\beta$ from the microscope images and substituting the values into the equations of motion for a rigid body given above, along with the experimentally extracted average velocity of a single particle $V_{JP}$ (which for an applied field of 134 V$_{RMS}$/cm is equal to $3.18 \mu m/s$ (Figure 4b)). The experimental values of $V_C$ (here denoted by $V_{C-EXP}$) were calculated using the instantaneous position of the centre $C$ in each frame, while $\omega$ was obtained by performing an FFT over the instantaneous orientation of the line connecting the two spheres relative to a fixed coordinate system.

We found that while the experimental and kinematic values for the translational velocity were similar, the model consistently over predicts $\omega$ by approximately 40% (see also Figure 4a below) and we thus chose to adjust the kinematic model to account for this reduction as a fitting parameter by letting $\omega_{KIN} = 0.6(\sin\alpha - \sin\beta)V_{JP}/2R$ (physical justification for this reduction will be discussed below).

It is immediately apparent that part d exhibits the maximum angular velocity and smallest translational velocity and radius of orbit. This result is to be expected when we note that the configuration of this doublet is closest to the "ideally aligned" case (i.e. $\alpha = 3\pi/2, \beta = \pi/2$) where the rigid body model predicts maximum angular velocity with zero net translation. In contrast, the velocity components perpendicular to the line of centers of the two spheres in Figure 3b, induce rotation in opposite directions so that the angular velocity is very small while the radius of orbit is rather large. Figure 3c lies in between these two extremes, although since both particles drive rotation in the same direction, the resultant path is closer to 3d than 3b.

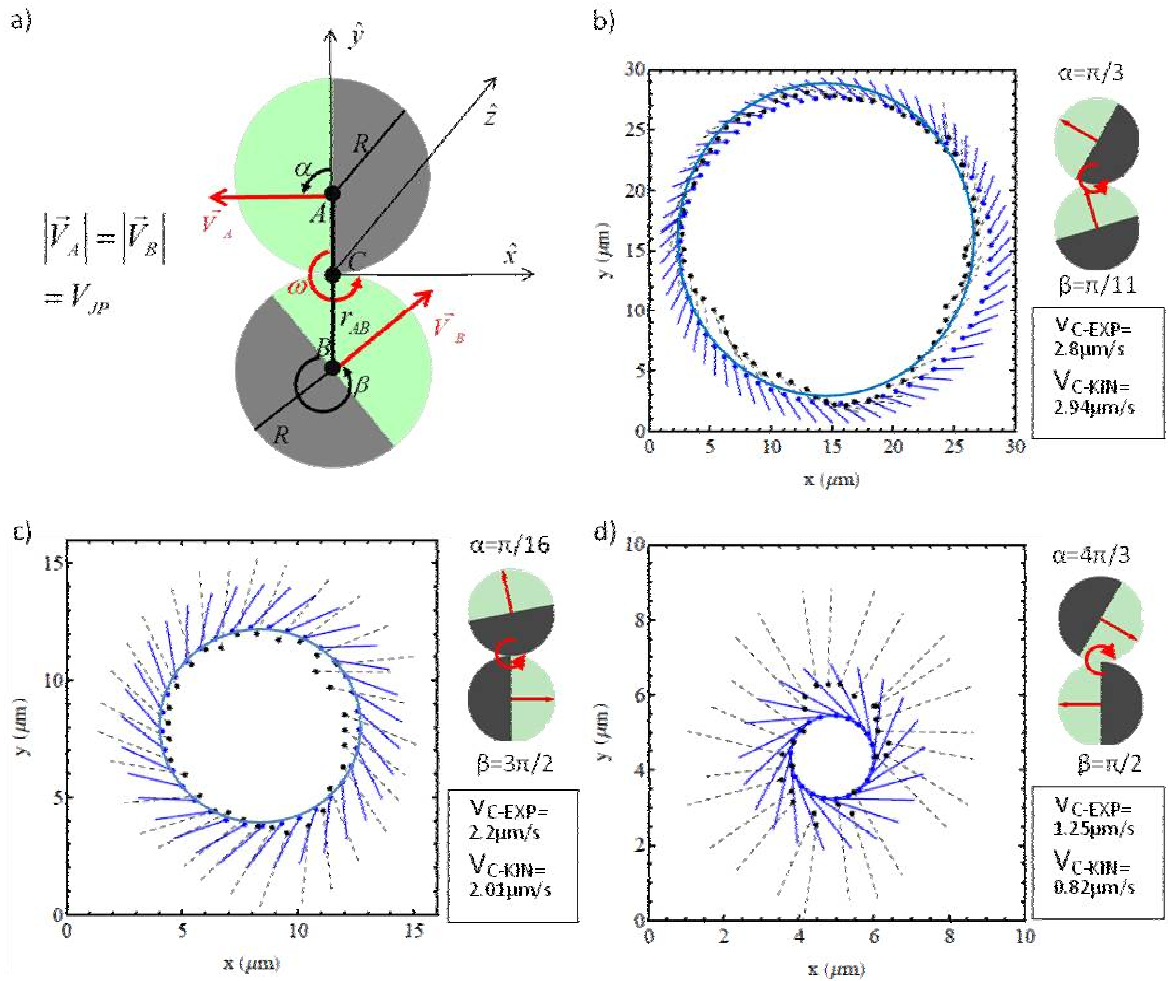

**Figure 3: a) Schematic of the rigid body model; b-d) Comparison of experimental results (dashed lines) with orbits predicted by the adjusted kinematic model (solid blue lines). Applied field is held constant at 134V/cm and 1.5kHz. The dark sides of the colloids correspond to the gold coated hemispheres.**

In both parts b and c of Figure 3, we observe that the experimental orbits are approximately circular and there is good agreement between the angles of the kinematic and experimental lines of centers. On the other hand, in 3d the orbit is somewhat elliptical resulting in a slight mismatch between the angles of the theoretical and experimental lines of centres. This discrepancy may be due to some 3D effects where the elliptical orbit indicates departure from planar motion and symbolizes the projection of an out-of-plane circular orbit onto the plane of view.

In Figure 4, we further examine the voltage dependent behavior of the doublets and plot the absolute values of the translational and angular velocities of a number of different Janus pairs with various interfacial alignments $(\alpha, \beta)$ as a function of the square of the applied field. Both the angular and linear velocities of the doublets are linearly proportional to $V^2$ in accordance with Eq.(1). This quadratic behavior is characteristic of non-linear electrokinetic flows and serves to distinguish them from their linear counterparts. However, as was also observed by Gangwal et al. [17], the velocities are an order of magnitude smaller than that predicted by the theory for a particle in an infinite medium (Eq.(1)). This discrepancy cannot be solely attributed to increased Stokes drag near a wall but rather confirms that the motion is not governed solely by induced-charge electrokinetic effects and that DEP resulting from non-uniformity of the field near the wall, and which tends to retard ICEP, is also a contributor [37] so that the motion is in fact "dipolophoretic" [40] in nature. Additionally, it has been shown that the relatively high applied voltages, which violate the 'weak field' assumption used in the theoretical derivation of (0), can cause a reduction in mobility [41].

In accordance with the rigid body model, we observe that those pairs which are close to "ideal alignment" (e.g., doublets 4,5 in Fig 4) exhibit maximum angular and minimum translational velocities. On the other hand, doublet 6, where $\alpha \square \beta$, has a minimum angular velocity and maximum translational velocity. In order to further compare our results with the rigid body model, we use a line of best fit through the experimental values of the translational velocity of a single particle (blue circles) to determine $V_{JP}(E)$ and then calculate the expected $\omega_{KIN}$ and $V_{C-KIN}$ for the three different combinations of $\alpha, \beta$. For the angular velocity, we include theoretical curves for both the unaltered and reduced (by 40% as in Figure 3) values of $\omega_{KIN}$.

From Figure 4a, we observe that the maximum translational velocity of a doublet approaches that of a single particle which is in agreement with the kinematic model. However, as was already discussed in Figure 3, the angular velocity (Figure 4b) consistently remains at about 60% of the value predicted by naively substituting the experimental value of $V_{JP}$ as the driving velocity.

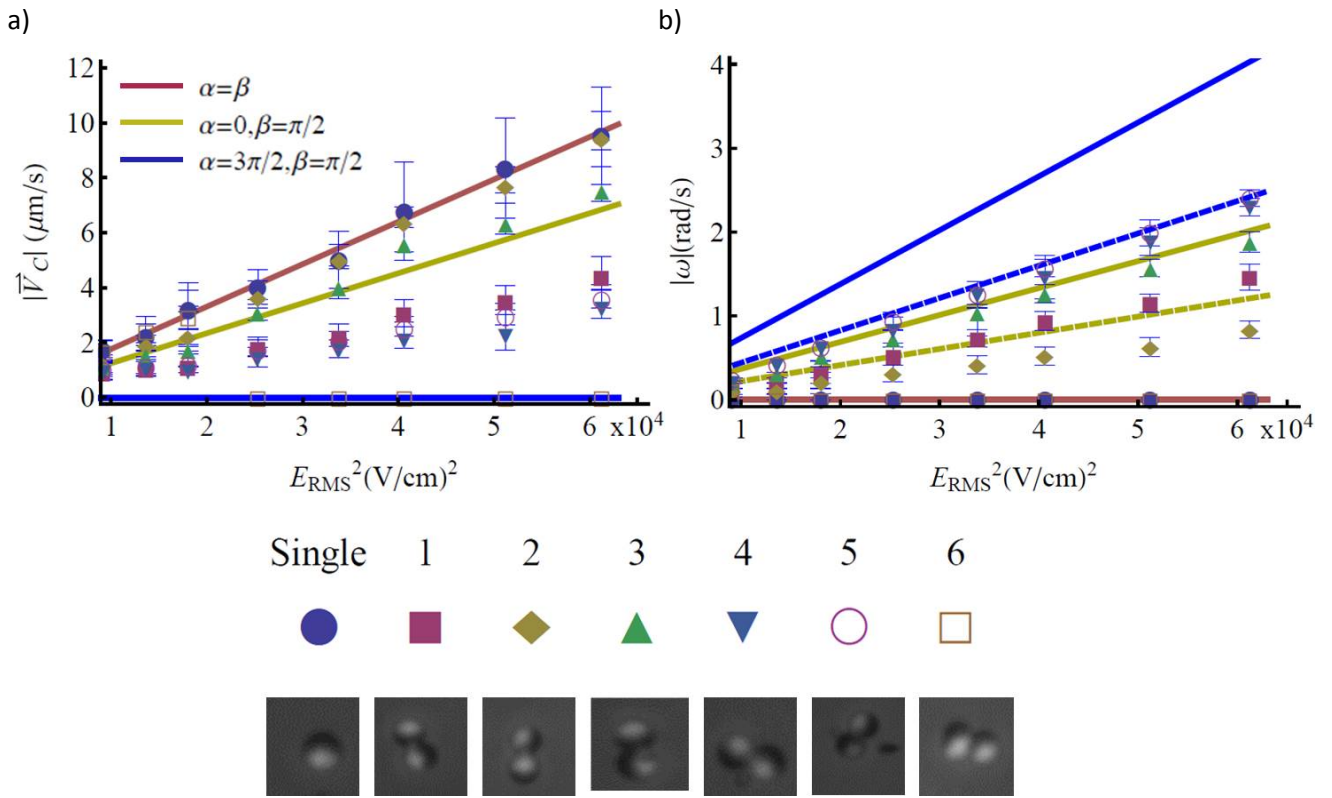

**Figure 4: Plots of a) linear velocity and b) angular velocity of multiple Janus pairs with varying metallodielectric interface orientations as a function of applied field squared. (Note that particles 2,3 and 4 correspond to Figures 2b,c,d and 3b,c,d respectively)**

To examine whether the discrepancies between the experimental and kinematic models could be the result of particle-particle interactions, we simulated the two extreme cases of $\alpha = 3\pi/2, \beta = \pi/2$ and $\alpha = \beta$ in Comsol$^{TM}$ (see supplementary materials [39]). As expected, the model predicts a reduction in mobility, which for the oversimplified simulated scenario (2D simulations) was ~40%. However, although this interference may explain the smaller experimental values of $\omega$, it also predicts a similar reduction in the values of $V_C$ which surprisingly is not evident in the experimental results. Thus it appears that the reduced rotation is more likely to stem from other sources such as error in the calculation of $\alpha, \beta$ or an unequal distribution of velocity between the two particles in each pair which effectively increases the misalignment. Such an imbalance could stem from manufacturing defects [25] and the fact that since the pairs are rigidly attached, the interfaces are not necessarily precisely aligned with the electric field which is expected to result in a reduction of mobility and 3D effects [21]. Additionally, it has been shown that the presence of any surface contamination or inhomogeneities can significantly reduce ICEO [42].

Nevertheless, it is evident that the kinematic model successfully predicts both the direction and relative magnitudes of the doublet velocities and their trajectories. Thus we demonstrate that despite the presence of presumed electrohydrodynamic particle-particle interactions, the primary determinant of the rotational properties of a pair of rigidly attached Janus particles are the relative angles of the interfaces between the conducting and dielectric hemispheres with respect to line of centers. We verify that both angular and linear velocities are quadratic in the applied field, consistent with non-linear electrokinetic effects. The application of a field perpendicular to the wall combined with the wall-attraction effect stabilizes rotation in the plane of view and facilitates examination of 3D symmetry breaking via the angles of the interfaces. Additionally, the proximity of the observed rotation to the wall suggests that while the motion is dominated by ICEP, the DEP contribution stemming from the non-uniformity of the electric field at the wall should be further investigated. To conclude, we emphasize that the present demonstration of broken-symmetry induced continuous rotation in a simple uniform AC field, aside from being a topic of fundamental interest, unlocks an additional degree of freedom, i.e., continuous rotation rather than linear translation, to the large number of micro and nanofluidic systems driven by such fields. Additionally, although the JP doublet constitutes the simplest geometry required to induce rotation via symmetry breaking, the approach herein may be generalized to more complex structures containing three or more JPs (e.g. [10]) and even composites of JP "carriers" and neutral "cargo".

**Acknowledgements**

The authors would like to acknowledge US-Israel Binational Science Foundation Grant 2009371. The JP preparation was possible through the financial and technical support of the Technion RBNI (Russell Berrie Nanotechnology Institute) and MNFU (Micro Nano Fabrication Unit). We also thank Israel Rozinsky and Yoav Green for technical assistance and Dr. Jarrod Schiffbauer for helpful discussions.

[1]  D. Long and A. Ajdari, Phys. Rev. Lett. **81**, 1529 (1998).
[2]  J. L. Anderson, J. Colloid Interface Sci. **105**, 45 (1985).
[3]  P. García-Sánchez, Y. Ren, J. J. Arcenegui, H. Morgan, and A. Ramos, Langmuir **28**, 13861 (2012).
[4]  J. J. Arcenegui, A. Ramos, P. García-Sánchez, and H. Morgan, ELECTROPHORESIS **34**, 979 (2013).
[5]  S. L. Biswal and A. P. Gast, Anal. Chem. **76**, 6448 (2004).
[6]  P. Garstecki, P. Tierno, D. B. Weibel, F. Sagués, and G. M. Whitesides, J. Phys. Condens. Matter **21**, 204110 (2009).
[7]  S. K. Smoukov, S. Gangwal, M. Marquez, and O. D. Velev, Soft Matter **5**, 1285 (2009).
[8]  V. A. Murtsovkin and G. I. Mantrov, Colloid J USSR **52**, 933 (1990).
[9]  E. Yariv, Phys. Fluids **17**, 051702 (2005).
[10] T. M. Squires and M. Z. Bazant, J. Fluid Mech. **560**, 65 (2006).
[11] I. Frankel, G. Yossifon, and T. Miloh, Phys. Fluids **24**, 012004 (2012).
[12] G. Yossifon, I. Frankel, and T. Miloh, Phys. Fluids **19**, 068105 (2007).
[13] K. A. Rose, J. A. Meier, G. M. Dougherty, and J. G. Santiago, Phys. Rev. E **75**, 011503 (2007).
[14] H. Sugioka, Phys. Rev. E **81**, 036301 (2010).
[15] S. Gangwal, O. J. Cayre, and O. D. Velev, Langmuir ACS J. Surfaces Colloids **24**, 13312 (2008).
[16] L. Zhang and Y. Zhu, Langmuir ACS J. Surfaces Colloids **28**, 13201 (2012).

[17] S. Gangwal, O. J. Cayre, M. Z. Bazant, and O. D. Velev, Phys. Rev. Lett. **100**, 058302 (2008).
[18] Y. Daghighi, I. Sinn, R. Kopelman, and D. Li, Electrochimica Acta **87**, 270 (2013).
[19] T. Honegger, O. Lecarme, K. Berton, and D. Peyrade, in (AVS, 2010), pp. C6I14–C6I19.
[20] T. Miloh, Phys. Fluids **20**, 107105 (2008).
[21] A. M. Boymelgreen and T. Miloh, ELECTROPHORESIS **33**, 870 (2012).
[22] W. Gao, A. Pei, X. Feng, C. Hennessy, and J. Wang, J. Am. Chem. Soc. **135**, 998 (2013).
[23] L. Baraban, M. Tasinkevych, M. N. Popescu, S. Sanchez, S. Dietrich, and O. G. Schmidt, Soft Matter **8**, 48 (2011).
[24] J. G. Gibbs, S. Kothari, D. Saintillan, and Y.-P. Zhao, Nano Lett. **11**, 2543 (2011).
[25] N. A. Marine, P. M. Wheat, J. Ault, and J. D. Posner, Phys. Rev. E **87**, 052305 (2013).
[26] L. Qin, M. J. Banholzer, X. Xu, L. Huang, and C. A. Mirkin, J. Am. Chem. Soc. **129**, 14870 (2007).
[27] G. Loget and A. Kuhn, Nat. Commun. **2**, 535 (2011).
[28] T. Mirkovic, N. S. Zacharia, G. D. Scholes, and G. A. Ozin, Small **6**, 159 (2010).
[29] J. Wang and W. Gao, ACS Nano **6**, 5745 (2012).
[30] P. Tierno, R. Albalat, and F. Sagués, Small **6**, 1749 (2010).
[31] L. Baraban, D. Makarov, O. G. Schmidt, G. Cuniberti, P. Leiderer, and A. Erbe, Nanoscale **5**, 1332 (2013).
[32] S. J. Ebbens and J. R. Howse, Soft Matter **6**, 726 (2010).
[33] M. García, J. Orozco, M. Guix, W. Gao, S. Sattayasamitsathit, A. Escarpa, A. Merkoçi, and J. Wang, Nanoscale **5**, 1325 (2013).
[34] A. Walther and A. H. E. Müller, Chem. Rev. **113**, 5194 (2013).
[35] M. B. Smith, E. Karatekin, A. Gohlke, H. Mizuno, N. Watanabe, and D. Vavylonis, Biophys. J. **101**, 1794 (2011).
[36] M. S. Kilic and M. Z. Bazant, ELECTROPHORESIS **32**, 614 (2011).
[37] T. Miloh, ELECTROPHORESIS **34**, 1939 (2013).
[38] S. Ebbens, R. A. L. Jones, A. J. Ryan, R. Golestanian, and J. R. Howse, Phys. Rev. E Stat. Nonlin. Soft Matter Phys. **82**, 015304 (2010).
[39] Supplementary Materials, (n.d.).
[40] V. N. Shilov and V. R. Estrela-Lopis, *Surface Forces in Thin Films and Dispersed Systems* (Consultants Bureau, New York, 1975).
[41] M. Z. Bazant, M. S. Kilic, B. D. Storey, and A. Ajdari, *Towards an Understanding of Induced-charge Electrokinetics at Large Applied Voltages in Concentrated Solutions* (2009).
[42] A. J. Pascall and T. M. Squires, Phys. Rev. Lett. **104**, 088301 (2010).